\documentclass[preprint,review,10pt,3p]{elsarticle}

\usepackage{amssymb}
\usepackage{amsthm}

\usepackage{lineno}

\journal{Physics Letters B}

\usepackage{graphicx}
\usepackage{amssymb}
\usepackage{dcolumn}
\usepackage{bm}
\usepackage{epsfig}
\usepackage{verbatim}
\usepackage{amsmath}
\usepackage{amsfonts}
\usepackage{epic}
\usepackage{eepic}
\usepackage{color}
\usepackage{perpage}
\biboptions{numbers,sort&compress}

\usepackage{url}
\definecolor{blue}{RGB}{0,128,178}
\usepackage[colorlinks=true,linkcolor=blue,urlcolor=blue,citecolor=blue]{hyperref}
\usepackage{breakurl}

\def\Edecay{50 $\pm$ 3~}
\def\Cexp{61 $\pm$ 6~}  
\def\Sexp{4.1 $\pm$ 0.4} 
\def\Cth{83.1~} 
\def\Csp{13.5~} 
\def\Rs{0.73 $\pm$ 0.07}

\begin{document}

\begin{frontmatter}


%
%
%

\title{Neutron occupancy of the $0d_{5/2}$ orbital and the $N$ = 16 shell closure in $^{24}$O}

\author[1]{K.~Tshoo\corref{cor1}}
\cortext[cor1]{Corresponding author.}
\ead{tshoo@ibs.re.kr}
%
\address[1]{Rare Isotope Science Project, Institute for Basic Science, Daejeon 305-811, Republic of Korea}
\author[2]{Y.~Satou}
\address[2]{Department of Physics and Astronomy, Seoul National University, Seoul 151-742, Republic of Korea}
\author[3]{C.A.~Bertulani}
\address[3]{Texas A$\&$M University-Commerce, PO Box 3011, Commerce, Texas 75429, USA}
\author[2]{H.~Bhang}
\author[2]{S.~Choi}
\author[4]{T.~Nakamura}
\author[4]{Y.~Kondo}
\author[4]{S.~Deguchi}
\author[4]{Y.~Kawada}
\author[4]{Y.~Nakayama}
\author[4]{K.N.~Tanaka}
\author[4]{N.~Tanaka}
\author[4]{Y.~Togano}
\address[4]{Department of Physics, Tokyo Institute of Technology, Tokyo 152-8551, Japan}
\author[5]{N.~Kobayashi}
\address[5]{Department of Physics, University of Tokyo, Tokyo 113-0033, Japan}
\author[6]{N.~Aoi} 
\address[6]{Research Center for Nuclear Physics, Osaka University, Osaka 567-0047, Japan}
\author[7]{M.~Ishihara}
\author[7]{T.~Motobayashi}
\author[7]{H.~Otsu}
\author[7]{H.~Sakurai}
\author[7]{S.~Takeuchi}
\author[7]{K.~Yoneda}
\address[7]{RIKEN Nishina Center, Saitama 351-0198, Japan}
\author[8]{F.~Delaunay}
\author[8]{J.~Gibelin}
\author[8]{F.M.~Marqu\'es}
\author[8]{N.A.~Orr}
\address[8]{LPC-Caen, ENSICAEN, Universit\'e de Caen, CNRS/IN2P3, 14050 Caen cedex, France}
\author[9]{T.~Honda}
\address[9]{Department of Physics, Rikkyo University, Tokyo 171-8501, Japan}
\author[10]{T.~Kobayashi}
\author[10]{T.~Sumikama}
\address[10]{Department of Physics, Tohoku University, Miyagi 980-8578, Japan}
\author[11]{Y.~Miyashita}
\author[11]{K.~Yoshinaga} 
\address[11]{Department of Physics, Tokyo University of Science, Chiba 278-8510, Japan}
\author[12]{M.~Matsushita}
\author[12]{S.~Shimoura}
\address[12]{Center for Nuclear Study, University of Tokyo, Saitama 351-0198, Japan}
\author[13]{D.~Sohler}
\address[13]{Institute for Nuclear Research of the Hungarian Academy of Sciences, PO Box 51, H-4001 Debrecen, Hungary}
\author[2]{J.W.~Hwang}
\author[14]{T.~Zheng}
\author[14]{Z.H.~Li}
\author[14]{Z.X.~Cao}
\address[14]{School of Physics and State Key Laboratory of Nuclear Physics and Technology, Peking University, Beijing 100871, China}


\begin{abstract}
One-neutron knockout from $^{24}$O leading to the first excited state in $^{23}$O has been measured for a proton target at a beam energy of 62 MeV/nucleon.~The decay energy spectrum of the neutron unbound state of $^{23}$O was reconstructed from the measured four momenta of the $^{22}$O fragment and emitted neutron.~A sharp peak was found at $E_{\rm decay}$ = \Edecay keV, corresponding to an excited state in $^{23}$O at 2.78 $\pm$ 0.11 MeV, as observed in previous measurements.~The longitudinal momentum distribution for this state was consistent with $d-$wave neutron knockout, providing support for a $J^\pi$ assignment of 5/2$^+$.~The associated spectroscopic factor was deduced to be $C^2S(0d_{5/2})$ = \Sexp~by comparing the measured cross section ($\sigma^{\rm exp}_{-1n}$ = \Cexp mb) with a distorted wave impulse approximation calculation.~Such a large occupancy for the neutron $0d_{5/2}$ orbital is in line with the $N$ = 16 shell closure in $^{24}$O.

\begin{flushleft}
$PACS:$ {24.50.+g, 21.10.Jx, 25.60.Gc, 27.30.+t}
\end{flushleft}
\end{abstract}


\end{frontmatter}


%


The neutron drip-line nucleus $^{24}$O, which has been studied extensively in recent years~\cite{Ozawa00,Rituparna01,Stanoiu04,BrownRichter,Alexander05,Hoffman08,Hoffman09,Kanungo09,Otsuka01,Otsuka05,Otsuka10,Jensen11,Hoff11,Tshoo2012}, is now considered a doubly-closed-shell nucleus.~In particular, the high excitation energy of the first $2^+$ state~\cite{Hoffman09} and small quadrupole transition parameter~\cite{Tshoo2012} are strong indicators of an $N$ = 16 spherical shell closure.~In addition, Kanungo et al.~\cite{Kanungo09} found a large spectroscopic factor $-$ $C^2S(1s_{1/2})$ = 1.74 $\pm$ 0.19 $-$ for one-neutron knockout from $^{24}$O reflecting an almost complete occupancy of the $1s_{1/2}$ orbital.

In this Letter we report on the spectroscopic factor for $d_{5/2}$ neutron removal from $^{24}$O.~The first excited state of $^{23}$O, which is neutron unbound~\cite{Schiller2007,Hoff11, satou_fwebody13}, was populated by one-neutron knockout from $^{24}$O with a proton target.~The excitation energy, cross section, and longitudinal momentum distribution were determined and allowed the $0d_{5/2}$ neutron-hole nature of this state to be identified.~The large spectroscopic factor deduced for this state is in line with the $N$ = 16 shell closure in $^{24}$O.



\begin{figure}[t]
\begin{center}
\resizebox{0.6\columnwidth}{!}{%
\includegraphics{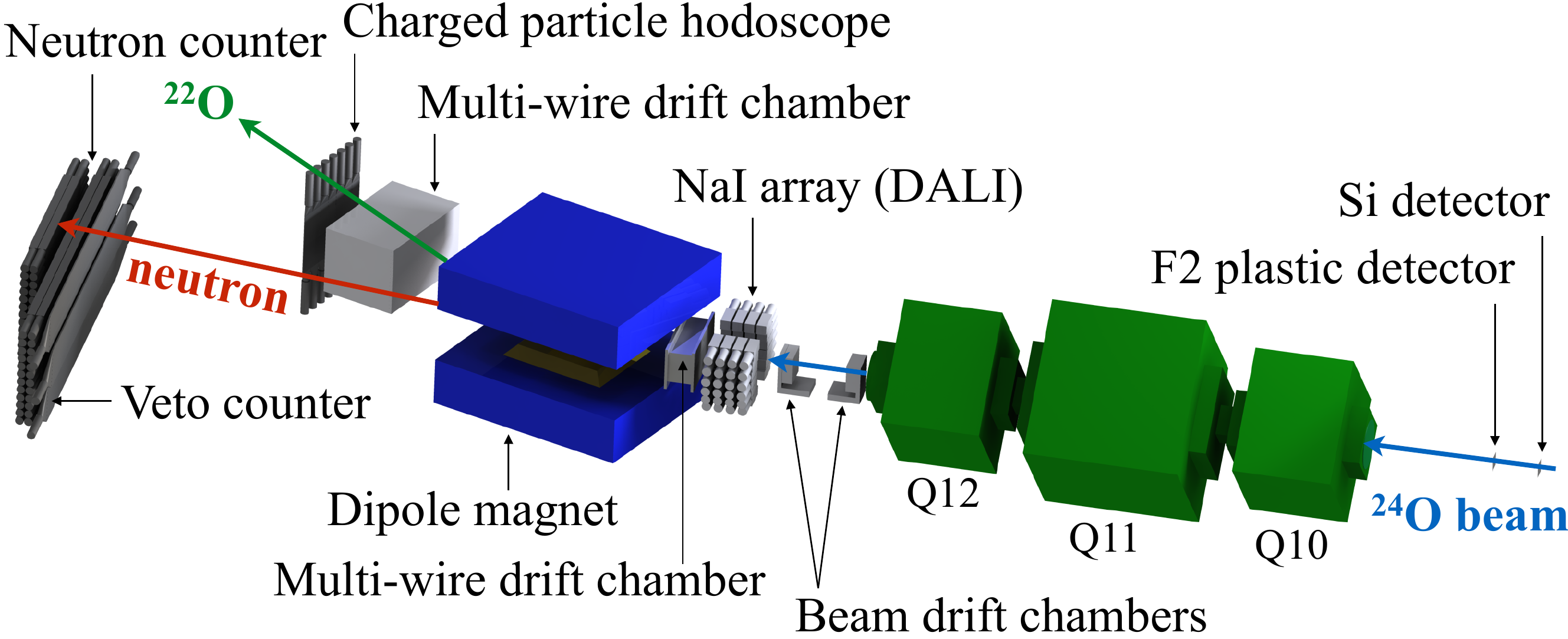}%
}
\caption{(Color online.) Schematic view of the experimental setup.
\label{fig1}}
 \end{center}
\end{figure}

The experiment was performed at the RIPS facility~\cite{Kubo} at RIKEN.~The experimental setup has been described in Refs.~\cite{Tshoo2012, satou2013PLB}, and is depicted in Fig.~\ref{fig1}.~The secondary $^{24}$O beam was produced using a 1.5 mm-thick Be production target and a 95 MeV/nucleon $^{40}$Ar primary beam of $\sim$40 pnA.~The average intensity of the $^{24}$O beam was $\sim$4 particles/sec.~The momentum of the secondary beam was determined particle-by-particle by measuring the position at the dispersive focus F1 of RIPS with a parallel plate avalanche counter.~The energy loss ($\Delta$$E$) and time-of-flight (TOF) were measured using 350 $\mu$m-thick silicon and 500 $\mu$m-thick plastic scintillator detectors, respectively, at the achromatic focus F2.~The liquid-hydrogen (LH$_{2}$) target~\cite{LH2} was installed at the achromatic focus F3.~The effective target thickness and the mid-target energy of $^{24}$O were 159 $\pm$ 3 mg/cm$^2$ and 62 MeV/nucleon, respectively.~The $^{24}$O beam incident on the target was tracked particle-by-particle by using two multi-wire drift chambers installed just upstream of the target.~The target was surrounded by an array consisting of 48 NaI(Tl) crystals (DALI) to detect $\gamma$ rays from de-excitation of the fragments.


The $B$$\rho$-TOF-$\Delta$$E$ method was employed to determine mass and charge of the fragment following reactions of the $^{24}$O beam with the LH$_2$ target.~The magnetic rigidity ($B\rho$) was determined from the position and angle information measured with two multi-wire drift chambers placed at the entrance and exit of the dipole magnet.~The TOF of the fragment was measured with the plastic scintillator charged particle hodoscope which also gave energy loss information.~The beam velocity decay neutron was detected using a plastic scintillator counter array placed some 4.7 m downstream of the target, equipped with a charged particle veto counter.~A neutron detection efficiency of 25 $\pm$ 1$\%$ at 64 MeV was measured for a 2 MeVee threshold in a separate $^7$Li($p,n$) measurement performed during the experiment.

\begin{figure}[t]
\begin{center}
\resizebox{0.6\columnwidth}{!}{%
\includegraphics{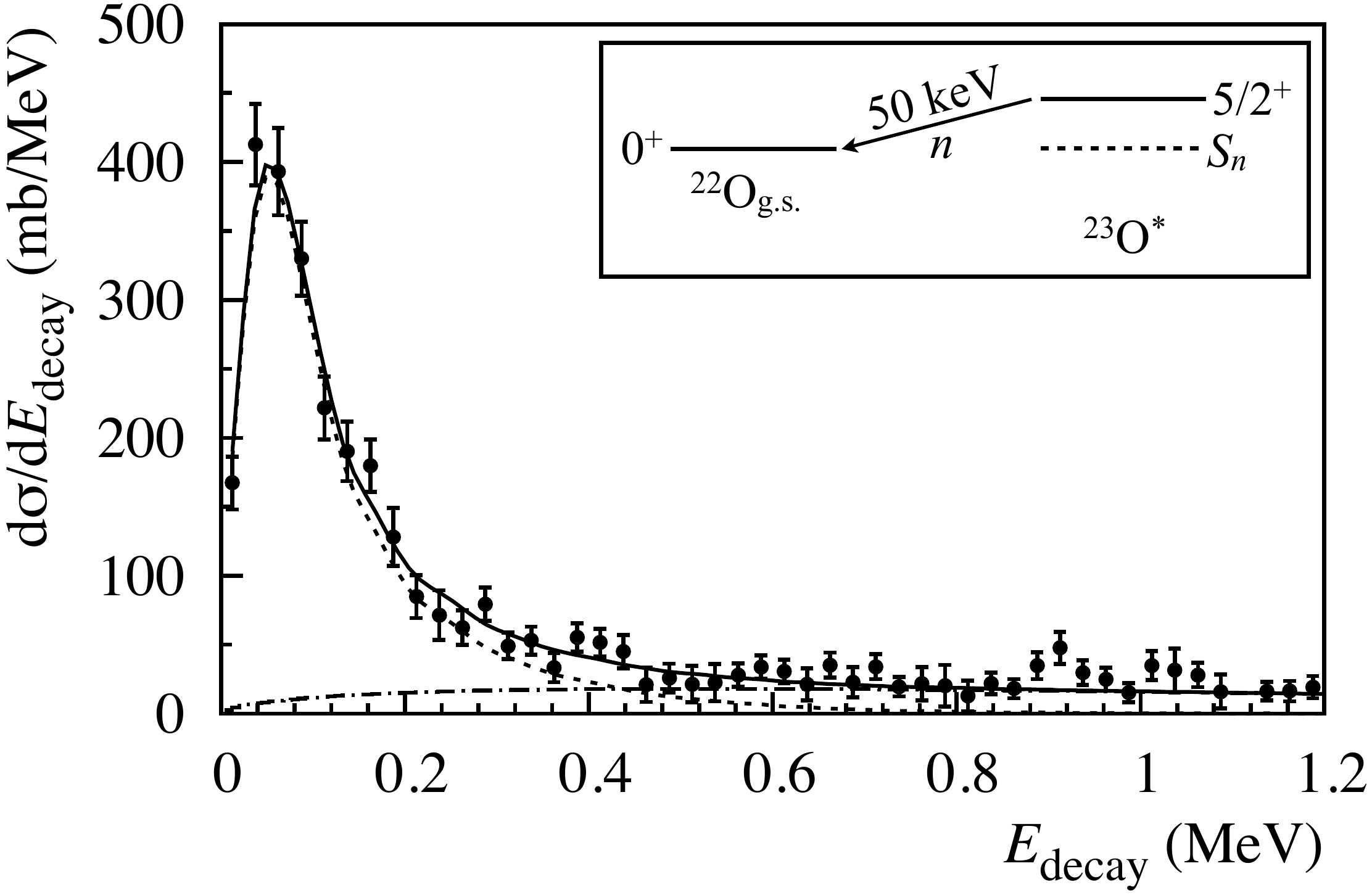}%
}
\caption{The cross section, d$\sigma$/d$E_{\rm decay}$, obtained by measuring a neutron in coincidence with $^{22}$O.~The error bars are statistical only.~The dotted and dot-dashed lines are the results of the fits for a resonance at $E_{\rm decay }$ =  \Edecay keV and a non-resonant continuum component~\cite{DEAK}, respectively.~The decay path from the first excited state of $^{23}$O to the ground state of $^{22}$O is represented in the inset. 
\label{fig2}}
 \end{center}
 \end{figure}

The decay energy spectrum of $^{23}$O$^*$ was reconstructed from the measured four momenta of $^{22}$O and the emitted neutron.~The decay energy $E_{\rm decay}$ is expressed as:
\begin{equation}
E_{\rm decay}=\sqrt{ (E_f + E_n)^2-|\boldsymbol{p}_f +\boldsymbol{p}_n|^2}-(M_f+M_n)~, 
\label{eq:proportionality_1}
\end{equation}
where $E_f~(E_n)$, $\boldsymbol{p}_f ~(\boldsymbol{p}_n)$, and $M_f~(M_n)$ are the total energy, momentum, and mass of $^{22}$O (neutron), respectively.~Fig.~\ref{fig2} shows the decay energy spectrum in terms of cross section (d$\sigma$/d$E_{\rm decay}$) after correcting for the detection efficiencies and acceptances.~The background contribution was subtracted here by using data taken with an empty target.~The error bars are statistical only.~The geometrical acceptance was estimated using a Monte Carlo simulation taking into account the beam profile, geometry of the setup, experimental resolutions, and multiple scattering of the charged particles.~The experimental energy resolution was estimated to be $\Delta E_{\rm decay}$ $\approx$ 0.52$\sqrt{E_{\rm decay}}$ (MeV) in FWHM.

The decay energy spectrum was described using one resonance for the peak at $E_{\rm decay}$ = \Edecay keV and a Maxwellian distribution for the non-resonant continuum~\cite{DEAK}.~The peak observed has an asymmetric lineshape owing to its proximity to threshold, and its width is fully dominated by the experimental resolution.~The corresponding excitation energy is $E_{\rm x}$ = 2.78 $\pm$ 0.11 MeV, given the separation energy of $S_{n}$($^{23}$O) = 2.73 $\pm$ 0.11 MeV~\cite{ame2012,Jurado,Orr_mass}.~Since no $\gamma$-ray lines were observed for the $^{22}$O$-$$n$ channel, we assumed that $^{22}$O is in the ground state.~As such, the location of the resonance is in good agreement with previous experiments~\cite{Schiller2007,satou_fwebody13}.

The one-neutron knockout cross section to the resonance at $E_{\rm decay}$ = 50 keV was determined to be $\sigma_{-1n}^{\rm exp}$ = \Cexp mb, after subtracting the non-resonant continuum.~The quoted error arises from the uncertainty in the choice of the functional form describing the non-resonant continuum (8$\%$), statistical uncertainty (4$\%$), neutron detection efficiency (3$\%$), and the target thickness (2$\%$).

The single-particle configurations for the ground states of $^{23,24}$O are expected to be $\nu$($0d_{5/2}$)$^{6}$$\nu$($1s_{1/2}$)$^1$ and $\nu$($0d_{5/2}$)$^{6}$$\nu$($1s_{1/2}$)$^2$, respectively, in the shell-model picture (see, e.g., Refs.~\cite{Kanungo09,Cortina04}), and suggest that the first excited state of $^{23}$O presented here (neutron hole state) is very likely to be populated by the $0d_{5/2}$ neutron knockout.~Calculations using the USDB interaction~\cite{USDab} in the $sd$ model space, as well as WBT~\cite{WBT} interaction in the $spsdpf$ model space, predict that the spin-parity of this state is $J^\pi = 5/2^+$ (Table~\ref{tab-1}).

The one-neutron knockout cross section, above a few tens MeV/nucleon energy, can be decomposed into the single-particle cross section ($\sigma_{\rm sp}$) related to the reaction and the spectroscopic factor ($C^2S$) reflecting the nucleon occupancy~\cite{Tostevin}, as follows:
\begin{equation}
\sigma_{-1n} = \sum_{nlj} \left( \frac{A}{A-1}\right)^\mathit{\Lambda} C^2S(J^\pi,nlj)~\sigma_{\rm sp}(nlj,S^{\rm eff}_n)~,
\label{eq2}
\end{equation}
where $J^\pi$ is the spin-parity of the final state of the residue (core);~$nls$ denote the quantum numbers of the knocked-out neutron;~$S^{\rm eff}_n$ is the effective separation energy which corresponds the sum of the neutron separation energy of the projectile and the excitation energy of the core;~$A$ is the mass number of the projectile;~$(A/(A-1))^\mathit{\Lambda}$ is the center-of-mass correction factor~\cite{Dieperink} where $A$ is the mass number of the projectile and $\mathit{\Lambda}$ is the major oscillator quantum number given by the relation of $\mathit{\Lambda}$ = $2n$ + $l$.

The single-particle cross section assuming a spectroscopic factor of unity was calculated using the distorted wave impulse approximation (DWIA) calculation for the quasifree ($p$,$pn$) reaction as described in Ref.~\cite{ppn_bertl}.~The calculation employed the eikonal approximation for the distorted wave functions of incoming and outgoing reaction channels.~The effective nucleon-nucleon interaction M3Y~\cite{M3Y} along with the Coulomb potential was adopted for the real part of the optical potential.~The nucleon-nucleus cross section in nuclear matter developed in Refs.~\cite{Bertulani10, Clementel, Bertulani86} and the densities of the projectile, or core, folded with the matter density of proton were employed to introduce the imaginary part of the optical potential.

\begin{figure}[t]
\begin{center}
\resizebox{0.6\columnwidth}{!}{%
\includegraphics{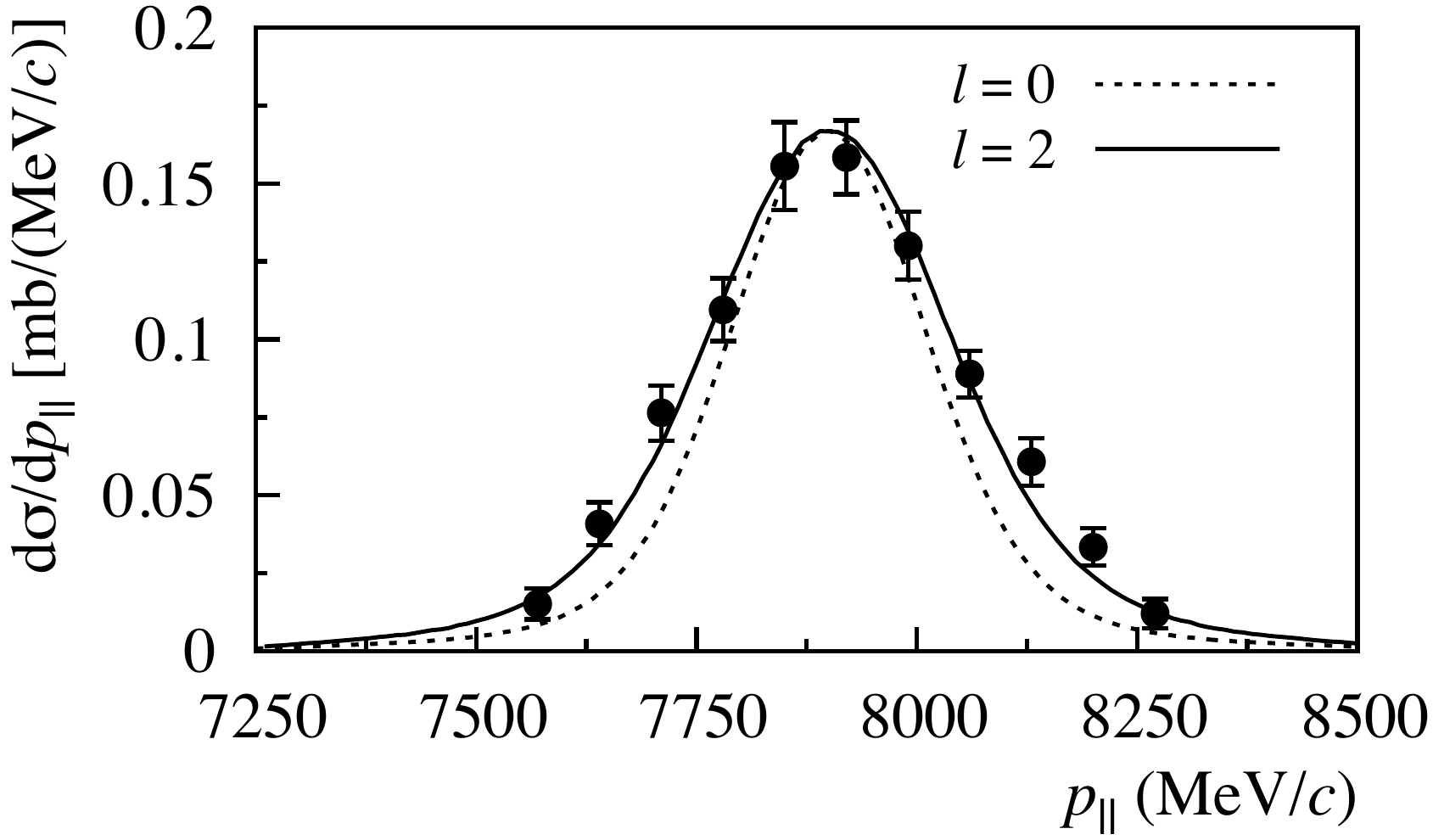}%
}
\caption{Longitudinal momentum distribution for $^{23}$O$^*$ (see text).~The error bars are statistical only.~The solid and dashed lines represent the calculated longitudinal momentum distributions normalized to the peak of the measured distribution for the knockout of neutrons in the $0d_{5/2}$ and $1s_{1/2}$ orbitals, respectively.~The distributions were convoluted with the experimental resolution.
\label{fig3}}
 \end{center}
\end{figure}

The wave functions of single-particle neutron orbits around the core were calculated using the Woods-Saxon potential in the manner as described in Ref.~\cite{Gade08}.~The Skyrme (SkX) Hartree-Fock (HF) calculation~\cite{SKX} was employed to deduce the root-mean-squared (rms) radius $r_{\rm HF}$ of each single-particle orbit in the projectile.~The code {\sc nushell}~\cite{Nushell} was used for this calculation.~The HF rms radius of $^{24}$O was $r_{\rm HF}$ = 3.434 fm for the $0d_{5/2}$ orbital.~The reduced radius $r_0$ was determined so that the calculated single-particle wave function within the potential well adopting a diffuseness of $a_0$ = 0.7 fm satisfies $r_{\rm sp}$ = $\sqrt{A/(A-1)}~r_{\rm HF}$ at the HF predicted binding energy.~The reduced radius was calculated to be 1.171 fm for the $0d_{5/2}$ orbital.~With this $r_0$, the depth of the potential well was further adjusted to reproduce $S_{n}^{\rm eff}$.~The $S_{n}^{\rm eff}$ was calculated for the 2.78-MeV state to be 6.97 MeV by adopting $S_n(^{24}{\rm O})$ = 4.19 $\pm$ 0.14 MeV~\cite{ame2012}.

The single-particle cross section was calculated to be $\sigma_{\rm sp}$ = \Csp mb, from which we deduced the spectroscopic factor $C^2S^{\rm exp}(0d_{5/2})$ = \Sexp~by using Eq.~(\ref{eq2}). The one-neutron knockout cross section was calculated to be $\sigma_{-1n}^{\rm th}$($0d_{5/2}$) = \Cth mb by employing the shell-model spectroscopic factor of $C^2S^{\rm th}$ = 5.67 obtained using the USDB interaction. The reduction factor defined as $R_s$ = $\sigma^{\rm exp}_{-1n}/\sigma^{\rm th}_{-1n}$ was derived to be $R_s$ = \Rs~which is consistent with the systematics in Refs.~\cite{Gade04,Gade08}. Moreover, a reduction factor of $R_s$ = 0.84 $\pm$ 0.10 may be deduced for inclusive one-neutron knockout at 51 MeV/nucleon from the neighboring $N$ = 14 closed sub-shell nucleus $^{22}$O~\cite{Sauvan}.


\begin{table}[t]
\centering
\caption{The excitation energies of the first two states in $^{23}$O and the spectroscopic factors for neutron removal from $^{24}$O are compared with the results of the USDB and WBT shell-model calculations.}
\label{tab-1}      
\vspace{5pt}
\begin{tabular}{p{0.5cm}p{1.3cm}p{0.6cm}p{0.6cm}p{0.001cm}p{1.3cm}p{0.6cm}p{0.6cm}}
\hline\noalign{\smallskip}
     & \multicolumn{3}{c}{\small Energy (MeV)}  & & \multicolumn{3}{c}{\small Spectroscopic factor $C^2S$} \\
    \cline{2-4} \cline{6-8}\noalign{\smallskip}
\small$J^\pi$  &\small Exp  &\small USDB   &\small WBT & &\small Exp &\small USDB  & \small WBT\\
\hline\noalign{\smallskip}
\small$1/2^+$  &\small 0.0   &\small 0.0  &\small  0.0 & &\small 1.74~$\pm$~0.19\footnote{From Ref.~\cite{Kanungo09}.}  &\small 1.81 &\small 1.73\\
\small $5/2^+$  &\small  2.78~$\pm$~0.11 &\small 2.59  &\small   2.72 & &\small  \Sexp  &\small  5.67  &\small  5.52
\\
\hline\noalign{\smallskip}
\end{tabular}
\begin{flushleft} \small $^1$From Ref.~\cite{Kanungo09}.\end{flushleft}
\end{table}

 In the core ($^{23}$O) + $n$ system description, the width of the longitudinal momentum ($p_{||}$) distribution of the core is directly linked to the single-particle wave function of the knocked-out neutron, and can be utilized to identify its orbital angular momentum ($l$).~We have reconstructed here the longitudinal momentum distribution of $^{23}$O$^*$ from the measured momenta of the $^{22}$O fragment and neutron.

The $p_{||}$ distribution for the first excited state of $^{23}$O in the laboratory frame is displayed in Fig.~\ref{fig3}.~The error bars are statistical only.~The width of the distribution was determined to be 356 $\pm$ 28 MeV/$c$ (FWHM) by a Gaussian fit.~The experimental distribution is compared in Fig.~\ref{fig3} with the DWIA calculations.~The results of the calculation for the removal of $0d_{5/2}$ (solid line) and $1s_{1/2}$ (dashed line) neutrons are shown in Fig.~\ref{fig3}, where the calculated distributions were convoluted with the momentum resolution.~The momentum resolution was determined to be 195 MeV/$c$ (FWHM) using the Monte Carlo simulation taking into account the momentum spread of the $^{24}$O beam, range difference of incoming and outgoing particles inside the LH$_2$ target, Coulomb multiple scattering effects, and the detector resolutions.~The widths of the calculated distributions for the $0d_{5/2}$ and $1s_{1/2}$ knocked-out neutrons were to be 334 and 270 MeV/$c$ (FWHM), respectively.~The result for $0d_{5/2}$ neutron knockout reproduces well the momentum distribution, supporting the spin-parity assignment of $J^{\pi}$ = 5/2$^+$.~It should be noted that this state has also been measured via two-proton and one-neutron removal~\cite{Schiller2007} and proton inelastic scattering~\cite{satou_fwebody13}.

%
%

\begin{figure}[t]
\begin{center}
\resizebox{0.6\columnwidth}{!}{%
\includegraphics{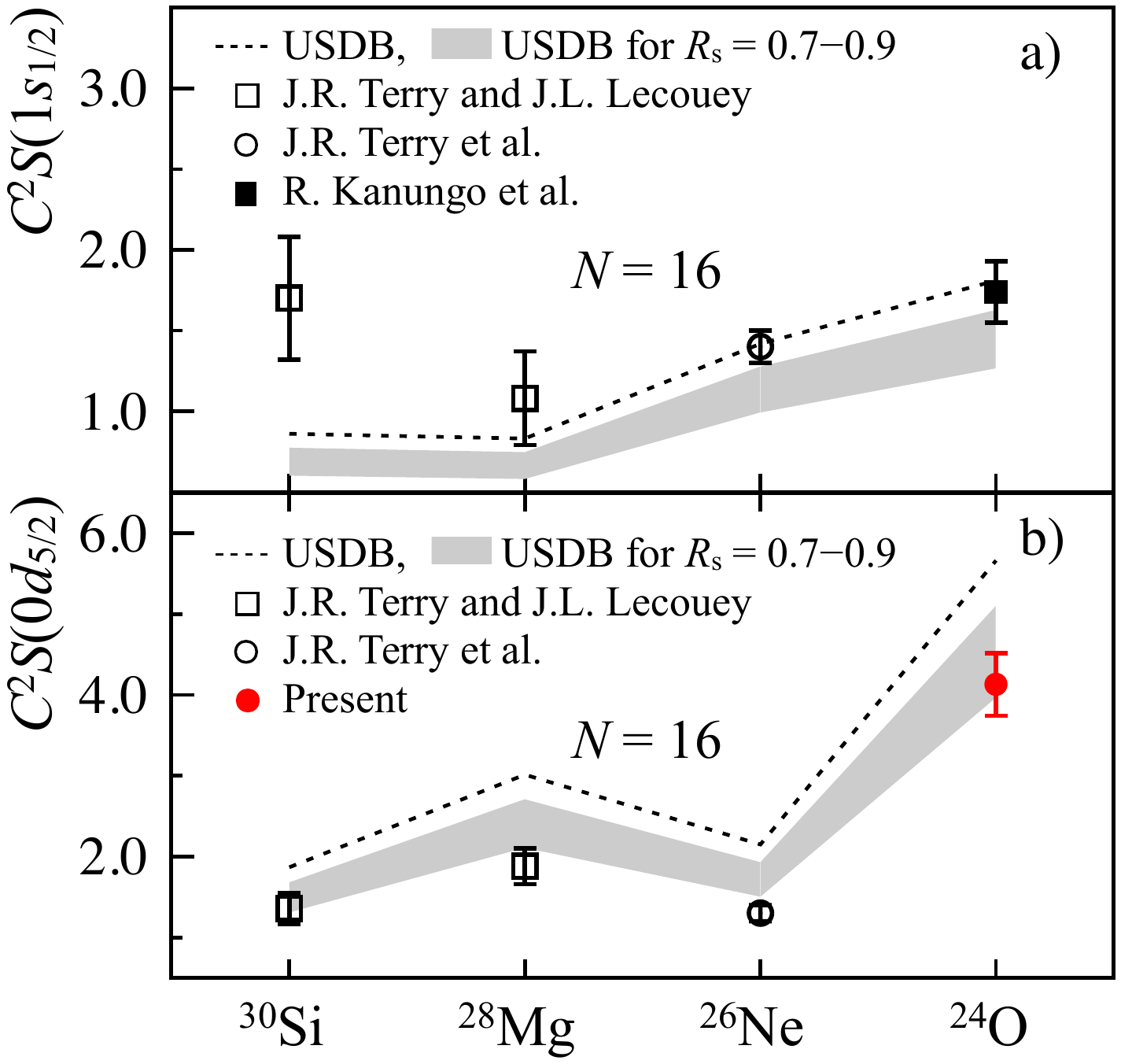}%
}
\caption{(Color online.) Spectroscopic factors of the lowest 1/2$^+$ and 5/2$^+$ states, $C^2S(1s_{1/2})$ (a) and $C^2S(0d_{5/2})$ (b), in even-even $N$ = 16 isotones.~The experimentally determined spectroscopic factors (data points) taken from Refs.~\cite{Kanungo09, TerryNPA04, Terry06} were derived from the neutron knockout reactions on a beryllium target.~The present result is shown by the red-filled circle.~The results of the USDB shell-model calculation are represented by the dashed lines.~The gray bands indicate the results of the calculations including a reduction factor $R_{\rm s}$, where the upper and lower limits of the bands correspond to 0.9 and 0.7, respectively.
\label{fig4}}
 \end{center}
\end{figure}

The excitation energy and the spectroscopic factor for neutron removal from $^{24}$O are compared with the results of the shell-model calculations using the USDB and WBT interactions in Table~\ref{tab-1}.~Both calculations reproduce the energy and also the spectroscopic factor assuming a reduction factor of $R_s$ $\approx$ 0.7$-$0.9 suggested by Gade et al.~\cite{Gade04,Gade08}.~We note that the USDB interaction successfully describes the energies of the first excited states and the quadrupole transition parameters of $^{22,24}$O reported in Refs.~\cite{Tshoo2012,Becheva06}.

Fig.~\ref{fig4} shows the proton number dependency of the experimentally determined and calculated (dashed lines) spectroscopic factors for the even-even $N$ = 16 isotones.~While the spectroscopic factor for the $\nu 1s_{1/2}$ orbital gradually increases in moving from $Z$ = 12 to 8, that for the $\nu 0d_{5/2}$ orbital dramatically rises at $Z$ = 8.~This trend is reasonably well reproduced by the USDB shell model calculations.~Significantly, the $C^2S(0d_{5/2})$ for $^{24}$O is much larger than those of other isotones, indicating a large neutron occupancy of the $0d_{5/2}$ orbital.~These results are in line with an $N$ = 16 sub-shell closure in $^{24}$O.

The gray bands in Fig.~\ref{fig4} represent the results of the USDB shell-model calculations including a reduction factor of $R_s$ = 0.7$-$0.9.~As such, the experimentally determined spectroscopic factors for the $\nu 0d_{5/2}$ orbit are in agreement with the calculations.~While the reduction factor is close to unity for the $1s_{1/2}$ neutron (for $Z$ $<$ 14), it is smaller than unity for the $0d_{5/2}$ neutron ($R_{\rm s}$ $\approx$ 0.7$-$0.8).~This behaviour may provide a benchmark to test more sophisticated nuclear structure models, as well as reaction mechanisms.~Recently, for example, Grinyer et~al.~\cite{Grinyer11} have investigated the quenching of the spectroscopic factors in $^{10}$Be and $^{10}$C for one-neutron knockout and found that an $ab~initio$ structure calculation, taking both 3-body forces and continuum effects into account, well describes the reduction.

In summary, we have investigated the occupancy of the neutron $0d_{5/2}$ orbital in $^{24}$O via one-neutron knockout from $^{24}$O with a proton target.~The excitation energy of the first excited state of $^{23}$O was measured to be 2.78 $\pm$ 0.11 MeV and the spin-parity was assigned, based on the longitudinal momentum distribution, to be $J^{\pi}$ = 5/2$^+$.~The large corresponding spectroscopic factor of $C^2S^{\rm exp}(0d_{5/2})$ = \Sexp~supports the picture of an $N$ = 16 spherical shell closure in $^{24}$O.

\section*{Acknowledgments}
We would like to thank the accelerator operations staff of RIKEN for providing the $^{40}$Ar beam.~This work is supported by the Grant-in-Aid for Scientific Research (No.~19740133) from MEXT Japan, the WCU program and Grant 2010-0024521 of the NRF Korea, the Rare Isotope Science Project of Institute for Basic Science funded by Ministry of Science and NRF of Korea (2013M7A1A1075765), DOE grants No. DE-FG02-08ER41533 and No. DE- FG02-10ER41706, and the French$-$Japanese LIA (IN2P3-RIKEN).





\end{document}